\documentclass[epj]{svjour}
\usepackage{english}
\usepackage{graphicx}

\newcommand{\la}{\langle}
\newcommand{\ra}{\rangle}
\newcommand{\bea}{\begin{eqnarray}}
\newcommand{\eea}{\end{eqnarray}}
\newcommand{\ltsim}{\mathop{\,<\kern-1.05em\lower1.ex\hbox{$\sim$}\,}}
\renewcommand{\arraystretch}{1.2}

\begin{document}

\title{Field-induced magnetic reorientation and effective anisotropy
of a ferromagnetic monolayer within spin wave theory}

\author{P. Fr\"obrich\inst{1,2} \and P.J. Jensen\inst{1,2}  
\and P.J. Kuntz\inst{1}}

\institute{Hahn-Meitner-Institut Berlin, Glienicker Stra{\ss}e 100, 
D-14109 Berlin, Germany \and 
also: Institut f\"ur Theoretische Physik, Freie Universit\"at Berlin\\
Arnimallee 14, D-14195 Berlin, Germany}

\date{Received: \today}

\titlerunning{Field-induced magnetic reorientation and effective
  anisotropy \ldots}

\authorrunning{P. Fr\"obrich \em et al.\em} 

\abstract{
The reorientation of the magnetization of a ferromagnetic monolayer
is calculated with the help of many-body Green's function theory.
This allows, in contrast to other spin wave theories, a satisfactory
calculation of magnetic properties over the entire temperature range of
interest since interactions between spin waves are taken into account. A
Heisenberg Hamiltonian plus a
second-order uniaxial single-ion anisotropy  and an external magnetic field is
treated by the Tyablikov (Random Phase Approximation: RPA) decoupling  of the
exchange interaction term and the Anderson-Callen decoupling   of the
anisotropy term. The orientation of the magnetization
is determined by the spin components $\la S^\alpha\ra$ ($\alpha=x,y,z$), 
which are calculated with the help of the spectral theorem. The knowledge 
of the orientation angle $\Theta_0$ allows a
non-perturbative determination of the temperature dependence of the
effective second-order anisotropy coefficient.
Results for the Green's function theory are compared with those
obtained with mean-field theory (MFT). We find significant differences between
these approaches.
}

\maketitle

\section{Introduction}\label{intro}
Experimental and theoretical investigations of the magnetic
properties of ultrathin ferromagnetic films are a topic of intense
current interest \cite{Hein}. In particular, the temperature dependent
(effective)
magnetic anisotropy and the resulting direction of the magnetization
have been determined for a variety of thin film systems \cite{exp}.
The different temperature and thickness dependence of the various
anisotropy contributions may result in a reorientation of the
magnetization as function of temperature and/or film thickness.
In addition, a magnetic reorientation can be induced by applying a
magnetic field \cite{dut,MOK98}, which is one of the experimental methods 
for measuring the magnetic anisotropies. In order to determine the different
contributions to the magnetic anisotropy separately, their dependence on the
temperature, on the film thickness, and on the magnetic field has to be known.

By use of \em ab-initio \em methods, the magnetic anisotropy of thin films
has been calculated as the total energy difference between
different directions of the magnetization \cite{anis}. Until now this has been
done only for $T=0$, whereas measurements are always performed at finite
temperatures. Thus, the knowledge of the temperature dependence
of the anisotropies is also required in order to allow a comparison of 
experimental and theoretical results.

This knowledge is gained mainly within the framework of a
Heisenberg model for localized spins, which, however, seems to yield
satisfactory results also for the important case of itinerant magnets such 
as Fe, Co, and Ni. Usually mean-field theory
(MFT) is applied to calculate the magnetization, accompanied either
by diagonalization of a single-particle Hamiltonian \cite{JeB98}, or by a
thermodynamic perturbation theory for the anisotropy terms \cite{theo}.
In principle, MFT is not applicable to two-dimensional (2D) magnetic systems.
In such systems, thermodynamic correlations, which are neglected by MFT, 
have a decisive influence on the magnetic properties.
In particular, the long-range magnetic fluctuations destroy the
remanent magnetization of an \em isotropic \em 2D Heisenberg model at
finite temperatures (Mermin-Wagner theorem \cite{MW66}). However, even small
anisotropic contributions, which are always present in real magnetic
systems, induce a magnetically ordered state in thin films with a
critical temperature of the order of the exchange coupling \cite{herr}. 
This is the reason why results obtained from MFT are expected to be
\em qualitatively \em correct also for 2D Heisenberg magnets.
Furthermore, it is known that applied magnetic fields have a much
larger impact on the magnetic properties of 2D 
than of three-dimensional (3D) systems \cite{yab}. Therefore, measured
magnetic quantities such as the effective anisotropy may in principle
depend on the experimental situation, e.g. whether a magnetic field is
present or not. The MFT method poorly reproduces this sensitive field
dependence of 2D systems, thus necessitating an improved theoretical
description.

Long-range magnetic fluctuations can be treated by spin wave 
theories. To our knowledge, the calculation of a field-induced
magnetic reorientation with such an approach has been performed only by
Erickson and Mills \cite{Mills}, who consider an exchange coupling,
a uniaxial lattice anisotropy, and the magnetic dipole coupling. They
have transformed the spin operators into Bose operators (Holstein-
Primakoff transformation \cite{HP40}), which are treat\-ed only in lowest 
order. Thus the validity of this linearized spin wave theory is limited
to low temperatures. The authors obtained a strong increase
of the transverse fluctuations near the reorientation transition,
where the direction of magnetization turns into the field
direction. However, as remarked by the authors, the
method should break down in this region.

In a previous paper \cite{EFJK99}, we demonstrated with the help of a 
many-body Green's function theory that the Tyablikov (or RPA) 
decoupling \cite{Tya67} of the higher-order Green's functions provides
a significantly improved description of the magnetization over MFT.
We showed this by comparing RPA and MFT results with the exact
solution of a Heisenberg spin pair, and with the `exact' Quantum
Monte Carlo result \cite{TGH98} of a Heisenberg monolayer with spin $S=1/2$.
Also the temperature dependence of the second- and fourth-order effective 
anisotropy coefficients was calculated by a thermodynamic
perturbation theory, expecting that RPA gives an improved description
also for the effective anisotropy coefficients. As results we found that their
temperature dependence looks different, particularly at low temperatures, 
and that their dependence on the magnitude of the spin is much weaker in 
RPA than in MFT \cite{EFJK99}. Such a perturbative approach makes sense 
only if the magnetic field is stronger than the anisotropy.

In the present work, we investigate also in the framework of many-body
Green's function theory the orientation of a ferromagnetic monolayer at finite
temperatures, the field-induced magnetic reorientation, and the effective 
(temperature-dependent) anisotropy. Encouraged by the fact that RPA yields
a good approximation to the magnetization, we expect that the RPA results 
of the present paper also represent more satisfactory estimates for the 
above mentioned magnetic properties than the results of MFT. 
Added to the Heisenberg Hamiltonian is the second-order
single-ion anisotropy favoring a perpendicular magnetization, and an external
magnetic field perpendicular to this uniaxial lattice anisotropy, causing 
a magnetic reorientation with an increasing field strength. Since we
are mainly interested in the action of the second order \em single-ion \em
anisotropy and its temperature dependence, we omit here the magnetic dipole
coupling. For a ferromagnetic monolayer the dipole coupling
competing with the out-of-plane second order single-ion anisotropy will induce
a reorientation as a function of the temperature only in a narrow parameter
range (for the strengths of the dipole coupling and the anisotropy) and
therefore a reorientation is quite improbable in general, see e.g.\
\cite{DIPmon}. The magnetization of a monolayer will in most cases
stay either in-plane or out-of plane. This is the reason why we study the
reorientation induced by an external magnetic field.
For several layers, on the other hand, the dipole coupling will play a more
important role. A magnetic reorientation as function of film thickness and/or
temperature can be expected, when the film surface and film interior 
anisotropic contributions compete with each other and exhibit different 
temperature dependences.

We apply the RPA method for decoupling the Green's functions coming from 
the exchange coupling terms thus approximately taking into account 
interactions between magnons, whereas we make use of the
Anderson-Callen decoupling \cite{AC64} for the corresponding anisotropy terms.
The magnetization axis will be tilted with respect
to the easy axis by the magnetic field; this procedure resembles the
experimental situation \cite{exp,dut}. We do not rotate the local spin
quantization axis but calculate the magnetization from the expectation values
of the spin components $\la S^\alpha \ra$. This gives the
equilibrium orientation angle directly.
Knowledge of the orientation angle allows a calculation of the
temperature dependent (effective) anisotropy coefficient from the
condition that the free energy be a minimum. This  is a non-perturbative
approach because the quantities entering in the final expression are 
calculated from the \em full \em Hamiltonian. This is an improvement over 
the previously \cite{EFJK99} used thermodynamic
perturbation theory, where the anisotropy term is treated as a small
perturbation. We compare the effective anisotropy obtained
from the non-perturbative approach with that from the thermodynamic
perturbation theory and, in many places throughout the paper,
we  compare the results obtained from the Green's function theory
with those from MFT.

\section{The Green's function formalism}\label{formalism}
Our aim is to determine the orientation angle $\Theta_0(T)$ of the
magnetization of a single (001)- layer as a function of the temperature $T$
from the expectation values $\la S^\alpha\ra$ ($\alpha=x,y,z$) 
of the components of the magnetization. For this purpose a Heisenberg 
Hamiltonian is used \cite{JeB98,theo} consisting of the isotropic exchange 
interaction $J$ between nearest neighbour lattice sites,
a second-order single-ion anisotropy parameter $K_{2}=K_2(T=0)$ at zero
temperature, and an external magnetic field, ${\bf{B}}=(B^x,B^y,B^z)$
\begin{eqnarray}
{\cal H}&=&-\frac{J}{2}\sum_{kl}{\bf{S}}_k{\bf{S}}_l-K_2\sum_k(S_k^z)^2
-\sum_k{\bf{B\,S}}_k \nonumber \\
&=&-\frac{J}{2}\sum_{kl}(S_k^-S_l^++S_k^zS_l^z)-K_2\sum_k(S_k^z)^2 
\nonumber \\
&&-\sum_k\Big(\frac{1}{2}B^-S_k^++\frac{1}{2}B^+S_k^-+B^zS_k^z\Big)\,, 
\label{1} \end{eqnarray}
where the notation
$S_i^\pm=S_i^x\pm iS_i^y$ and $B^\pm=B^x\pm iB^y$ is introduced.

In order to treat the problem, one needs the following Green's functions:
\begin{equation}
G_{ij(\eta)}^{\alpha,mn}=\la\la
S_i^\alpha;(S_j^z)^m(S_j^-)^n\ra\ra\;;\hspace{0.5cm}\alpha=+,-,z \,, 
\label{2} \end{equation}
where $\eta=\pm1$ refer to the commutator ($\eta=-1$)
or anti-commutator ($\eta=1$) Green's functions, respectively,
$n\ge1$ and $m\ge0$ are positive integers, $i$ and $j$ denote lattice sites.

The $G_{ij(\eta)}^{\alpha,mn}$ are determined from the equations of
motion in the spectral representation
\begin{equation}
\omega\;G_{ij(\eta)}^{\alpha,mn}(\omega)=A_{ij(\eta)}^{\alpha,mn}+\la\la
[S_i^\alpha,{\cal H}]_{-1};(S_j^z)^m(S_j^-)^n\ra\ra_\omega\,, 
\label{3} \end{equation}
with the inhomogeneities
\bea A_{ij(\eta)}^{\alpha,mn}&=&\la[S_i^\alpha,(S_j^z)^m(S_j^-)^n]
_{\eta}\ra \nonumber \\
&=&\la S_i^\alpha(S_j^z)^m(S_j^-)^n+\eta(S_j^z)^m(S_j^-)^nS_i^\alpha\ra\,, 
\label{4} \eea
where $\la...\ra={\rm Tr}(...e^{-\beta\cal{H}})$
with $\beta=1/k_BT$ and $k_B$ Boltzmann's constant.

Knowledge of the Green's functions allows the determination of the respective
correlation functions by the spectral theorem \cite{Tya67}
\bea 
C_{ij}^{\alpha mn}&=&\la (S_j^z)^m(S_j^-)^nS_i^\alpha\ra=
\frac{i}{2\pi}\lim_{\delta\rightarrow 0}\int_{-\infty}^{\infty}
\frac{d\omega}{e^{\beta\omega}+\eta} \times \nonumber \\ 
&& \bigg[G_{ij(\eta)}^{\alpha,mn}
(\omega+i\delta)-G_{ij(\eta)}^{\alpha,mn}(\omega-i\delta)\bigg]\,.
\label{5} \eea 
Calculation of the commutators $[S_i^\alpha,H]_{-1}$ yields the
following set of equations of motion for the Green's functions
\begin{eqnarray}
\omega G_{ij(\eta)}^{\pm,mn}&=&A_{ij(\eta)}^{\pm,mn}\mp J\sum_k
\big(\la\la S_i^zS_k^\pm;(S_j^z)^m(S_j^-)^n\ra\ra \nonumber \\ 
&& -\la\la S_k^zS_i^\pm;(S_j^z)^m(S_j^-)^n\ra\ra\big) \nonumber \\ 
&&\pm K_{2}\la\la(S_i^\pm S_i^z+S_i^zS_i^\pm);(S_j^z)^m(S_j^-)^n\ra\ra
\nonumber \\ 
&& \mp B^\pm G_{ij(\eta)}^{z,mn} \pm B^zG_{ij(\eta)}^{\pm,mn}\nonumber\\
\omega G_{ij(\eta)}^{z,mn}&=&A_{ij(\eta)}^{z,mn}+\frac{J}{2}\sum_k
\la\la(S_i^-S_k^+-S_k^-S_i^+); \nonumber \\
(S_j^z)^m&&\hspace{-0.5cm}(S_j^-)^n\ra\ra-\frac{1}{2}B^-G_{ij(\eta)}^{+,mn}
+\frac{1}{2}B^+G_{ij(\eta)}^{-,mn} \label{6}
\end{eqnarray}
The higher-order Green's functions occuring on the right-hand sides
have to be decoupled in order to obtain a closed set of equations.
For the exchange coupling terms we apply a generalized Tyablikov-
(or RPA-) \cite{Tya67,Tya59} decoupling, allowing also for a finite
value of the $x,y$- (or $\pm$-) components of the magnetization
($\alpha,\beta=+,-,z$; $i\neq k$)
\begin{equation}
\la\la S_i^\alpha S_k^\beta;(S_j^z)^m(S_j^-)^n\ra\ra\simeq
\la S_i^\alpha\ra G_{kj}^{\beta,mn}+\la S_k^\beta\ra
G_{ij}^{\alpha,mn}\,. \label{7}
\end{equation}
The terms resulting from the single-ion anisotropy ($i=k$) have to be
decoupled differently. A  RPA-decoupling, as was proposed by Narath
\cite{Na65}, is reasonable for an exchange anisotropy
($\propto S^z_iS^z_k$). It yields for the single-ion anisotropy
unphysical results, for instance for $S=1/2$ the respective terms do not
vanish. Instead, an ansatz of the following form is introduced \cite{Lin67}
\begin{equation}
\la\la S_i^\pm S_i^z+S_i^zS_i^\pm;(S_j^z)^m(S_j^-)^n\ra\ra 
\simeq \Phi_i^{\pm,mn} G_{ij}^{\pm,mn}, \label{7a}
\end{equation}
where the functions $\Phi_i^{\pm,mn}$ have to be determined. This type
of decoupling is valid for anisotropies which are small compared to the
exchange interaction. In Appendix A, we have investigated various 
decoupling schemes for the single-ion anisotropy proposed in the literature.
In the appendix, we give arguments for our preferring the Anderson-Callen 
decoupling for the present calculations, yielding the decoupling function
for $n=1$
\begin{equation} \Phi\equiv \Phi_i^{\pm,m1}
\simeq 2\la S_i^z\ra \left(1-\frac{1}{2S^2}[S(S+1)-\la
S_i^zS_i^z\ra]\right)\,. \label{8}
\end{equation}

We now apply the Tyablikov decoupling to the exchange interaction term and
the Anderson-Callen decoupling to the single-ion anisotropy term in equations
(\ref{6}). Performing in addition a 2D- Fourier transformation to momentum
space, with ${\bf k}=(k_x,k_y,0)$ being the in-plane wave vector, yields the
following set of equations of motion
\begin{equation}
\left( \begin{array}{ccc}
\omega-\tilde{H^z} & 0 & H^+ \\
 0 & \omega+\tilde{H^z} & -H^- \\
\frac{1}{2}H^- & -\frac{1}{2}H^+ & \omega \end{array} \right) \!\!
\left( \begin{array}{ccc}
G_{\eta}^{+,mn}({\bf{k}},\omega) \\ G_{\eta}^{-,mn}({\bf{k}},\omega)  \\
G_{\eta}^{z,mn}({\bf{k}},\omega) \end{array} \right) =
\left( \begin{array}{ccc} A_\eta^{+,mn} \\ A_\eta^{-,mn} \\
A_\eta^{z,mn} \end{array} \right) \;, \label{9}
\end{equation}
with the abbreviations
\begin{eqnarray}
H^\alpha&=&B^\alpha+\la S^\alpha\ra J(q-\gamma_{\bf k})\,,
\qquad \alpha=+,-,z \nonumber\\
\tilde{H}^z&=&H^z+K_2\,\Phi\;=\;Z+ \la S^z\ra J(q-\gamma_{\bf k})
\nonumber\\ Z&=&B^z+K_2\,\Phi\,. \label{10}
\end{eqnarray}
For a square lattice, one obtains $\gamma_{\bf k}=2(\cos k_x+\cos k_y)$,
and $q=4$ is the number of nearest neighbours.
Note that $A_\eta^{\alpha,mn}=A_\eta^{\alpha,mn}({\bf k})$ depends
on the wave vector ${\bf k}$ for $\eta=1$ but not for $\eta=-1$.

The determinant of the matrix in equation (\ref{9}) is given by
\begin{equation}
\Delta(\omega,{\bf{k}})=\omega(\omega-E_{\bf k})(\omega+E_{\bf k})\,, 
\label{10a} \end{equation}
which is the magnon dispersion relation 
\begin{equation}
E_{\bf k}=\sqrt{H^+H^-+\tilde{H^z}\tilde{H^z}}\,. \label{10b}
\end{equation}
Hence, the eigenvalues are
\begin{equation}
 \omega_1=0\;,\ \ \omega_{2,3}=\pm E_{\bf k}\,. \label{11}
\end{equation}
The Green's functions are given by
\begin{equation}
G_\eta^{\alpha,mn}(\omega,{\bf{k}})=\frac{\Delta_\eta^{\alpha,mn}
(\omega,{\bf{k}})}{\Delta(\omega,\bf{k})}\,, \hspace{0.5cm} \alpha=+,-,z\,,
\label{12} \end{equation}
where $\Delta_\eta^{\alpha,mn}$ is the determinant where column $\alpha$  of
$\Delta$ is replaced by the right-hand side of equation (\ref{9}).
One obtains, for example
\bea \Delta_\eta^{z,mn}(\omega,{\bf k})&=&A_\eta^{z,mn}\Big(\omega^2-
(\tilde{H^z})^2 \Big) \nonumber \\ 
-\frac{1}{2}A_\eta^{+,mn}&&\hspace{-0.5cm} H^-(\omega+\tilde{H^z})
-\frac{1}{2}A_\eta^{-,mn} H^+(\omega-\tilde{H^z})\,. \qquad \label{13}
\eea
Now we use the fact that the \em commutator \em Green's functions must
be regular for $\omega\rightarrow 0$, e.g. \cite{Tya67},
\begin{equation}
\lim_{\omega\rightarrow 0}\;\omega G_{-1}^{\alpha,mn}({\omega,\bf{k}})=0 \,.
\label{14} \end{equation}
Thus $\Delta_{-1}^{\alpha,mn}(0,{\bf k})=0$.
Since one of the eigenvalues vanishes, see equation (\ref{11}), we
obtain from equation (\ref{13}) for $\tilde{H^z}\neq0$
\begin{equation}
H^-A_{-1}^{+,mn}+H^+A_{-1}^{-,mn}+2\,\tilde{H^z}A_{-1}^{z,m n} = 0 \,.
\label{15} \end{equation}

Evaluating this expression for $m=0$ and $n=1$ we find together with the
definitions in equation (\ref{10})
\begin{equation}
\frac{H^\pm}{\tilde{H^z}}=\frac{B^\pm}{B^z+K_2\,\Phi}=\frac{B^\pm}{Z}\,.
\label{16} \end{equation}
Putting this into equation (\ref{15}) we have
\begin{equation}
-2Z\,A_{-1}^{z,mn}=A_{-1}^{+,mn}B^-+A_{-1}^{-,mn}B^+\,. \label{17}
\end{equation}
Equations (\ref{15}) and (\ref{17}) are important relations between 
correlation functions which we call the \em regularity conditions. \em 
We note that the same relations are obtained for all three $\alpha=+,-,z$.
With $m=0$ and $n=1$ we obtain a relation between
$\la S^\pm\ra$ and $\la S^z\ra$
\bea \la S^\pm\ra&=&\frac{B^\pm}{Z}\la S^z\ra \label{18} \\
&=&\frac{B^\pm\;\la S^z\ra}{B^z+2K_2\la
S^z\ra\left(1-[S(S+1)-\la S^zS^z\ra]/2S^2\right)}\,. \nonumber \eea
This means that the knowledge of $\la S^z\ra$ and $\la S^zS^z\ra$ also 
determines the expectation value of the spin components $\la S^x\ra$ and 
$\la S^y\ra$. Other 
useful relations obtained from equation (\ref{17}) are given in Appendix B.

It remains to establish equations which determine the moments 
$\la(S^z)^m\ra$. For this purpose we consider for example the
following commutator Green's function
\begin{eqnarray}
G_{-1}^{z,mn}(\omega,{\bf k})&=&\frac{\Delta_{-1}^{z,mn}(\omega,\bf{k})}
{\Delta(\omega,\bf{k})}
=\frac{1}{\omega(\omega-E_{\bf k})(\omega+E_{\bf k})} \times \nonumber \\
\bigg(-A_{-1}^{+,mn}&&\hspace{-0.5cm}\frac{1}{2}H^-\omega+A_{-1}^{-,mn}
\frac{1}{2}H^+\omega+A_{-1}^{z,mn}\omega^2\bigg) \,, \label{19}
\end{eqnarray}
where the regularity condition has been taken into account.

In order to calculate correlation functions in the case of a vanishing
eigenvalue one also needs the respective \em anti-commutator \em
Green's function $G_{+1}^{z,mn}$, since in this case the correct form of
the spectral theorem reads, e.g.\ \cite{Tya67}
\bea C_{\bf k}^{zmn}&=&
\frac{i}{2\pi}\lim_{\delta\rightarrow 0}\int_{-\infty}^{\infty}
\frac{d\omega}{e^{\beta\omega}-1} \times \nonumber \\
\Big[G_{-1}^{z,mn}(\omega\,&&\hspace{-0.45cm}+i\delta,{\bf k})
-G_{-1}^{z,mn}(\omega-i\delta,{\bf k})\Big] +D_{\bf k}^{z,mn}\,, \label{20}
\eea
where
\begin{equation}
D_{\bf k}^{z,mn}=\lim_{\omega\rightarrow 0}\;\frac{\omega}{2}\;
G_{+1}^{z,mn}(\omega,{\bf k})\,. \label{21}
\end{equation}
Using
\begin{eqnarray}
& & G_{+1}^{z,mn}(\omega,{\bf k})=
\frac{\Delta_{+1}^{z,mn}(\omega,{\bf k})}{\Delta(\omega,{\bf k})}
=\frac{1/2}{\omega(\omega-E_{\bf k})(\omega+ E_{\bf k} )} \nonumber \\
&\times&\bigg(-A_{+1}^{+,mn}\;H^-(\omega+\tilde{H^z})+
A_{+1}^{-,mn}\;H^+(\omega-\tilde{H^z}) \nonumber \\
&+&2\,A_{+1}^{z,mn}\Big(\omega^2-(\tilde{H^z})^2\Big)\bigg)\,, \label{22}
\end{eqnarray}
and the relation between anti-commutator and commutator correlation functions
\begin{equation}
A_{+1}^{\alpha,mn}({\bf k})=A_{-1}^{\alpha,mn}+2\,C_{\bf k}^{\alpha mn}\,, 
\label{23} \end{equation}
we find together with the regularity condition
\begin{equation}
D_{\bf k}^{z,mn}=\frac{1}{2E_{\bf k}^2}\bigg(
C_{\bf k}^{+mn}H^-\tilde{H^z}+C_{\bf k}^{-mn}H^+\tilde{H^z}
+2C_{\bf k}^{zmn}(\tilde{H^z})^2\bigg)\,. \label{24}
\end{equation}
Finally, we obtain from equation (\ref{20}) together with equations
(\ref{19}) and (\ref{24})
\begin{eqnarray}
&&2H^+H^-C_{\bf k}^{zmn}-H^-\tilde{H^z}C_{\bf k}^{+mn}-
H^+\tilde{H^z}C_{\bf k}^{-mn}= \nonumber \\ 
&&\frac{1}{2}A_{-1}^{+,mn}E_{\bf k}\,H^-\left[\frac{E_{\bf k}}
{\tilde{H^z}}-\coth(\beta E_{\bf k}/2)\right] \nonumber \\
&+&\frac{1}{2}A_{-1}^{-,mn}E_{\bf k}\,H^+\left[\frac{E_{\bf k}}{\tilde{H^z}}
+\coth(\beta E_{\bf k}/2)\right] \,. \label{25} 
\end{eqnarray}
From these relations and the regularity conditions, equation (\ref{17}), one
obtains all necessary expectation values.

In the following we restrict ourselves to an external magnetic field ${\bf B}$
confined to the $xz$-plane. Because of the azimuthal symmetry in the case 
of an uniaxial anisotropy it is sufficient to deal with the
$z$- and $x$- components of the magnetization ($\la S^y\ra=0$ for 
$B^y=0$). From these values the magnitude and the equilibrium polar angle
of the magnetization are determined:
\begin{eqnarray}
M^2(T)&=&\la S^x\ra^2+\la S^z\ra^2\,, \nonumber \\
\Theta_0(T)&=&\arctan\frac{\la S^x\ra}{\la S^z\ra}
=\arctan\frac{B^x}{B^z+K_2\Phi}\,. \label{27}
\end{eqnarray}

The knowledge of $M(T)$ and $\Theta_0$ enables a non-pertur\-bative
determination of the temperature dependence of the anisotropy coefficient.
The anisotropic part of the free energy is usually written as a power 
series of the direction cosines of the magnetization. Its precise form is 
written in accordance with the symmetry 
of the system and usually converges. This series need not only be
valid for small fields and anisotropies as compared to the exchange coupling.
It can even be considered as a definition of the 
effective (temperature-dependent) anisotropy coefficients. It is such an
expression from which experimentalists determine the effective anisotropies.
The corresponding part of the free energy in the presence of the lowest 
order term, the second-order uniaxial anisotropy $K_2(T)$, and the Zeeman 
term has the form
\begin{equation}
F(T,\Theta)=F_0(T)-K_2(T)\cos^2\Theta-{\bf{B\,\cdot M(T)}}\,. \label{28}
\end{equation}
This corresponds to the physical situation where the
higher order anisotropies can be neglected. We only treat this case in the
present paper. $K_2(T)$ is then determined from the condition that the 
free energy has a minimum at the equilibrium angle $\Theta_0$: 
\bea \frac{\partial F(\Theta)}{\partial\Theta}(\Theta_0)&=&0=
2\,K_2(T)\cos\Theta_0\sin\Theta_0 \nonumber \\ 
&-&B^xM(T)\cos\Theta_0+B^zM(T)\sin\Theta_0\,. \qquad \label{29} \eea
Solving with respect to $K_2(T)$ we find
\begin{equation}
K_2(T)=\frac{M(T)(B^x\cos\Theta_0-B^z\sin\Theta_0)}
{2\cos\Theta_0\sin\Theta_0}\,. \label{30} \end{equation}
Note that the effective anisotropy can be considered as an intrinsic 
property of the layer itself only if $K_2(T)$ as obtained by equation 
(\ref{30}) is practically independent of the external magnetic field.
We mention that the effective anisotropy at zero temperature
$K_2(T=0)$ differs from the $K_2$ in the Hamiltonian by a spin-dependent
normalization factor \cite{JeB98,MF95}. By expanding equations (\ref{8}),
(\ref{27}) and (\ref{30}) for small $T$ and $\Theta_0$ one finds
$K_2(T=0)=K_2\ S(S-1/2)$. The results shown in figures 3, 4, 6 are normalized
with respect to this factor for $S=1$.

The procedure described above is non-perturbative in the sense that the
quantities (the magnetization and the 
orientation angle) entering in equation (\ref{30}) are calculated from
the \em full \em Hamiltonian. This is in contrast to a thermodynamic
perturbation theory in which the lattice anisotro\-py is used as a small 
perturbation. The Hamiltonian is split into two terms 
${\cal H}={\cal H}_0-K_2\sum_l(S_l^z)^2$, and the effective 
anisotropy is calculated from moments of the unperturbed Hamiltonian 
${\cal H}_0$ only, see e.g.\ Ref.\ (\cite{EFJK99}).

In the present thermodynamic approach $K_2(T)$
depends on the temperature mainly
through the magnetization, which itself is a function of the magnetic field.
Thus at least a weak field dependence of the effective anisotropy is expected.
There are other sources which might induce a temperature dependence. In Ref.
{\cite{MHB96}} it has been shown within a tight-binding scheme that at a 
\em constant \em magnetization, the electronic entropy, the thermal lattice 
expansion, the population change of spin-orbit-splitted energy levels near 
the Fermi energy cause a decrease in the uniaxial anisotropy of a freestanding
Fe monolayer at a temperature  of about 1000 K. However, the magnetization 
and thus the resulting effective anisotropy will already have vanished
below 500 K for this case. Therefore the decreasing magnetization is the main
source for the temperature dependence of $K_2(T)$ in the temperature range we
are looking at in the present paper.

In the following, we treat as an example $S=1$, which is the lowest spin
value with a nontrivial second-order anisotropy.
For this case, the single-ion decoupling function reads
$\Phi=\la S^z\ra \la S^zS^z\ra$, see equation (\ref{8}).
Then we obtain from equation (\ref{25}) with $n=1,m=0$ and $n=1,m=1$, 
respectively, two coupled equations for $\la S^z\ra$ and
$\la S^zS^z\ra$, after taking into account the regularity 
conditions of Appendix B for $S=1$. The resulting equations are
\bea \la S^zS^z\ra&-&\frac{2}{1+(B^x/Z)^2} \nonumber \\
&+&\frac{\la S^z\ra}{2}\;\frac{2-(B^x/Z)^2}{\sqrt{1+(B^x/Z)^2}}\;\phi(T)=0 \,, 
\label{26b} \eea \vspace{-0.3cm}
\bea &&\la S^z\ra\Big(1-(B^x/Z)^2\Big)\Big(2-(B^x/Z)^2\Big) 
+2\la S^zS^z\ra \times \nonumber \\
&&\Big(1+(B^x/Z)^2\Big) -\bigg[
2\Big(3\la S^zS^z\ra-2\Big)\Big(1-(B^xZ)^2\Big)\rule{1cm}{0cm} \nonumber \\ 
&&-\la S^z\ra\Big(2-(B^x/Z)^2\Big)\bigg]
\sqrt{1+(B^x/Z)^2}\;\;\phi(T)=4 \,, \label{26c} 
\end{eqnarray}
with
\begin{equation}
\phi(T)=\frac{1}{\pi^2}\int_0^\pi dk_x\int_0^\pi dk_y
\coth{(\beta E_{\bf k}/2)} \,.
\label{26a} \end{equation}
These equations have to be solved numerically in order to obtain
$\la S^z\ra$ and $\la S^zS^z\ra$. Together with
equation (\ref{18}) these determine the magnitude $M(T)$ and the orientation
angle $\Theta_0$ of the magnetization, cf. equations (\ref{27}).

In the next Section we compare the RPA method with a (Bragg-Williams) mean
field approximation for the exchange coupling term for a quantum-mechanical 
spin $S=1$. The resulting expectation values are obtained
by diagonalizing the corresponding dynamical matrix consisting of the 
molecular field, the external magnetic field and the single-ion anisotropy. 
For details see Ref.\ (\cite{JeB98}).
\begin{figure}[t] \label{Fig1}
\hspace*{2cm} 
\includegraphics[width=7.5cm,height=13cm,angle=180]{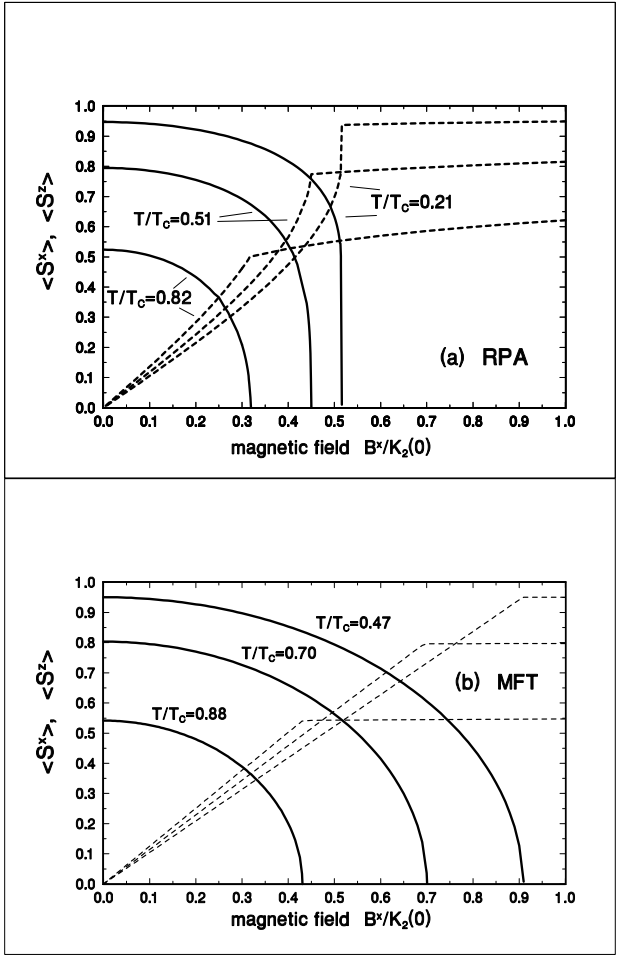}
\caption{Components of the magnetization $\la S^z\ra$ (solid lines)
and $\la S^x\ra$ (dashed lines) for different reduced
temperatures $T/T_C$ as  functions of the external magnetic field in
x-direction, 
$B^x/K_2(0)$, which is normalized to the anisotropy coefficient at temperature
$T=0$. Results of RPA (a) and MFT (b) calculations are compared.
The Curie temperatures are $T_C^{RPA}/J=0.989$ within RPA,
and $T_C^{MFT}/J=2.667$ within MFT, using $J/K_2(0)=100$. The reduced
temperatures in (a) and (b) are chosen in such a way that the magnetizations
$\la S^z\ra$ are approximately the same in RPA and MFT at $B^x=0$.
}\end{figure}

\section{Results} \label{Results}
In this section, we display results of our calculations for a Heisenberg 
Hamiltonian plus second-order uniaxial single-ion  anisotropy for a
square monolayer with spin $S=1$. If not stated otherwise, the 
interactions will be normalized to the single-ion anisotropy coefficient 
at zero temperature $K_2=K_2(T=0)>0$. We use for the
exchange coupling $J/K_2(0)=100$. With zero magnetic
field and these parameters, the RPA method predicts a Curie temperature
$T_C^{RPA}/J=0.989$, cf.\ Appendix A. The mean-field theory yields with
the same parameters a Curie temperature $T_C^{MFT}/J=2.667$, which is about a
factor of three larger. The temperature at which the magnetization 
reaches the field direction ($\la S^z\ra\to 0$) is
called the \em reorientation temperature \em $T_R$, and the
corresponding magnetic field the \em reorientation field \em $B^x_R$.

In Fig.1(a) we show the results obtained from the RPA method
for the components of the magnetization, $\la S^z\ra$ and $\la S^x\ra$, 
as functions of the external magnetic field
$B^x/K_2(0)$ in x-direction ($B^y=B^z=0$), for different reduced temperatures
$T/T_C$. This field-induced  magnetic reorientation is characterized by a
decreasing $\la S^z\ra$ and an increasing $\la S^x\ra$.
The magnetization reaches the in-plane direction ($\la S^z\ra=0$) at a field
strength $B^x_R$ depending on the temperature.
For the lowest temperature ($T/T_C=0.21$), we observe a  jump in
the components of the magnetization at the corresponding reorientation
field of about $B^x/K_2(0)\simeq 0.52$. This is probably caused by
the particular kind of the single-ion anisotropy decoupling applied in our
procedure. After complete reorientation $\la S^x\ra$ shows a 
nearly constant behaviour with increasing field.
In Fig.1(b) we show the corresponding MFT results with the same 
parameters for $J$ and $K_2(0)$. In order to compare the different
shapes of the magnetization curves in RPA and MFT, we have chosen
different reduced temperatures $T/T_C^{MFT}$
for the MFT calculations in such a way that the relative magnetizations 
$\la S^z\ra$ at $B^x=0$ are about the same in both
cases. One observes that $\la S^z\ra$
decreases near $B^x_R$ more rapidly within the RPA approach than
within MFT. Also, whereas $\la S^x\ra$ as computed with MFT
exhibits an almost linear behaviour as a function of $B^x/K_2(0)$, the
corresponding RPA- dependence is clearly curved.
The main difference is that the reorientation fields $B_R^x$
are considerably smaller within the RPA approach for all temperatures
investigated. This we attribute to the fact that a magnetic field has 
a stronger influence on 2D than on 3D systems. Note that the MFT results 
depend only on the coordination number and not on the spatial 
dimensionality, thus MFT handles a 2D system similar to a 3D one. 

In Fig.2(a,b) we display the orientation angles $\Theta_0(T,$ $B^x)$ 
corresponding to the situation of Fig.1 from a perpendicular
($\Theta_0=0^\circ$) to an 
in-plane direction ($\Theta_0=90^\circ$) of the monolayer magnetization.
\begin{figure}[t] \label{Fig2}
\vspace*{-0.8cm}\hspace*{-2cm} 
\includegraphics[width=9cm,height=13cm]{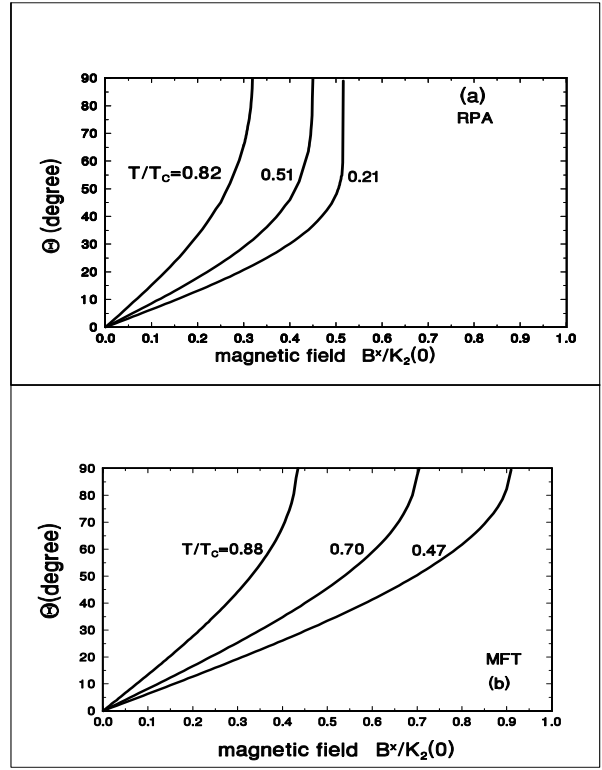}
\vspace{-1cm}

\caption{
The orientation angle $\Theta_0$ of the magnetization is shown as a 
function of $B^x/K_2(0)$ for the same reduced temperatures as in Fig.1. 
RPA results (a) are compared with MFT results (b).
}\end{figure}

The effective anisotropy coefficient, $K_2(T)$, is determined
by equation (\ref{30}), which we
apply for $B^z=0$. As mentioned before, this ansatz is physically 
meaningful only if the dependence of $K_2(T)$ on $B^x$ is small. In
Fig.3 the corresponding dependence is shown for different temperatures.
For all temperatures we obtain only a weak dependence of $K_2(T)$ on
$B^x/K_2(0)$ for small fields, which becomes stronger as the
reorientation field strength  $B^x_R$ is approached.
Therefore, we have used the small value $B^x/K_2(0)=0.1$  to
determine $K_2(T)$ as a function of the reduced temperature.
In this case the reorientation temperature $T_R$ is close to $T_C$.
The resulting effective anisotropy $K_2(T)/K_2(0)$ and the corresponding
orientation
angle $\Theta_0(T)$ are shown in Fig.4. $K_2(T)$ as obtained from RPA
is an almost straight line between $T=0$ and $T=T_R\simeq T_C$. On the other
hand, the corresponding behaviour of $K_2(T)$ calculated by MFT is linear
at elevated temperatures but shows an exponential behaviour when approaching
$T=0$. This has the consequence that for $S=1$ the MFT approach
yields a considerably smaller value of $K_2(0)$ than the RPA
method, if observed values of $K_2(T)$ (measured
e.g.\ at $T/T_C\sim0.7$ \cite{exp}) are extrapolated to $T=0$. Only at $T=0$
are anisotropy constants from ab-initio calculations available which can then
be compared with extrapolated measurements.
Note that the range of the exponential behaviour of $K_2(T)$ when applying 
MFT \cite{JeB98,theo} shrinks for large $S$ or for classical spins .
\begin{figure}[t] \label{Fig3}
\vspace*{-1.6cm}\hspace*{-1cm} 
\includegraphics[width=6cm,height=10cm,angle=-90]{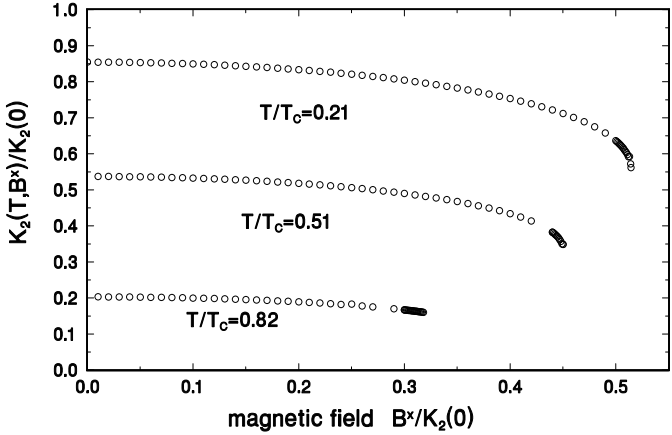}
\vspace{2cm}

\caption{Non-perturbative RPA-calculations for the effective single-ion 
anisotropy coefficient
$K_2(T,B^x)$ normalized to $K_2(0)$ are shown as a function of the external
magnetic field $B^x/K_2(0)$ for different reduced temperatures $T/T_C$.
}\end{figure}

We do not observe a large difference between RPA and MFT results for the 
orientation angle $\Theta_0(T)$ as a function of the reduced 
temperature $T/T_C$. Note, however, the different temperature scale. 
$T_C$ as obtained from MFT is 2.7 times larger
than the corresponding RPA result for the parameters under consideration.

In Fig.4 the reorientation temperature is close to the Curie temperature, 
since we have used a small field of $B^x/K_2(0)=0.1$. When applying 
a larger field, for instance $B^x/K_2(0)=0.4$, the reorientation takes 
place at lower temperatures. This is demonstrated in Fig.5(a),\break which
shows $\la S^z\ra$ and $\la S^x\ra$ as functions of the reduced temperature
resulting from RPA calculations, yielding a reorientation temperature
of about $T_R\simeq 0.65\,T_C$. The corresponding magnitude of the
magnetization $M(T)=\sqrt{\la S^x\ra^2+\la S^z\ra^2}$ is also shown, as 
well as the second moments $\la S^zS^z\ra$ and $\la S^xS^x\ra$, which 
approach the value $S(S+1)/3= 2/3$ for large temperatures. 
For comparison the corresponding results of a MFT calculation are
displayed in Fig.5(b). For the same applied field the reorientation 
temperature $T_R\simeq 0.9\,T_C$ is considerably higher in this case.
We emphasize the long tail in particular of the magnetization
$M(T)$ at large temperatures
in the RPA calculations of Fig.5(a), which is absent in the MFT results of
Fig.5(b). This behaviour is due to the strong effect of external magnetic
fields on the properties of 2D Heisenberg magnets \cite{yab}.
\begin{figure}[t] \label{Fig4}
\vspace*{-1.6cm}\hspace*{-1cm} 
\includegraphics[width=6cm,height=10cm,angle=-90]{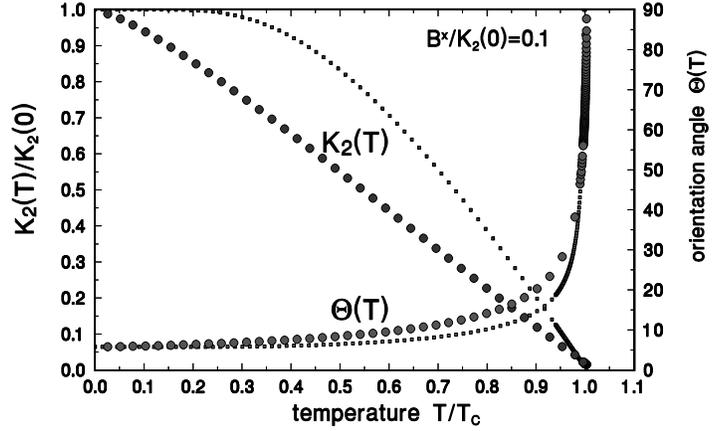}
\vspace{2cm}

\caption{Effective anisotropy coefficient $K_2(T)/K_2(0)$ at 
$B^x/K_2(0)=0.1$ and orientation angle
$\Theta_0(T)$ are shown as functions of the reduced temperatures
$T/T_{C}$. Large dots correspond to RPA, and small squares to MFT
results. }\end{figure}

In Ref.\ \cite{EFJK99} we showed the temperature dependence of
the single-ion anisotropy coefficients of a ferromagnetic monolayer
obtained with a thermodynamic perturbation theory.
The magnetization was calculated in the framework of RPA,
considering the isotropic exchange coupling and an external magnetic
field $B^z$ in the unperturbed Hamiltonian. In perturbation theory, it is
unavoidable to apply a finite magnetic field in order
to obtain a magnetization at finite temperatures (Mermin- Wagner theorem).
Thus, the perturbative results for the effective anisotropy $K_2(T)$
depend sensitively on the strength of the magnetic field $B^z$.
Even with this approach, we found significant differences from the 
corresponding mean-field theory calculations. As discussed in Section 
2, equation (\ref{30}) determines within RPA the effective single-ion 
anisotropy $K_2(T)$ non-perturbatively. In Fig.6 
we compare the non-perturbative result for $K_2(T)$, already shown in
Fig.4, with perturbative results as a function of the reduced temperature.
We use $B^z/K_2(0)=0.01$, $0.1$, and $1.0$. As already
mentioned the absolute value of
$K_2(T)/K_2(0)$ as calculated perturbatively depends considerably on the
magnetic field $B^z$. Except for the rounding at elevated temperatures,
which is clearly an effect of $B^z$, the shapes of $K_2(T)$
as calculated perturbatively and non-perturbatively within the RPA approach
look similar to each other. The differences between perturbative and 
non-perturbative results become larger with increasing $K_2(0)$.
\begin{figure}[t] \label{Fig5}
\vspace*{0cm}\hspace*{2cm} 
\includegraphics[width=7.5cm,height=13cm,angle=180]{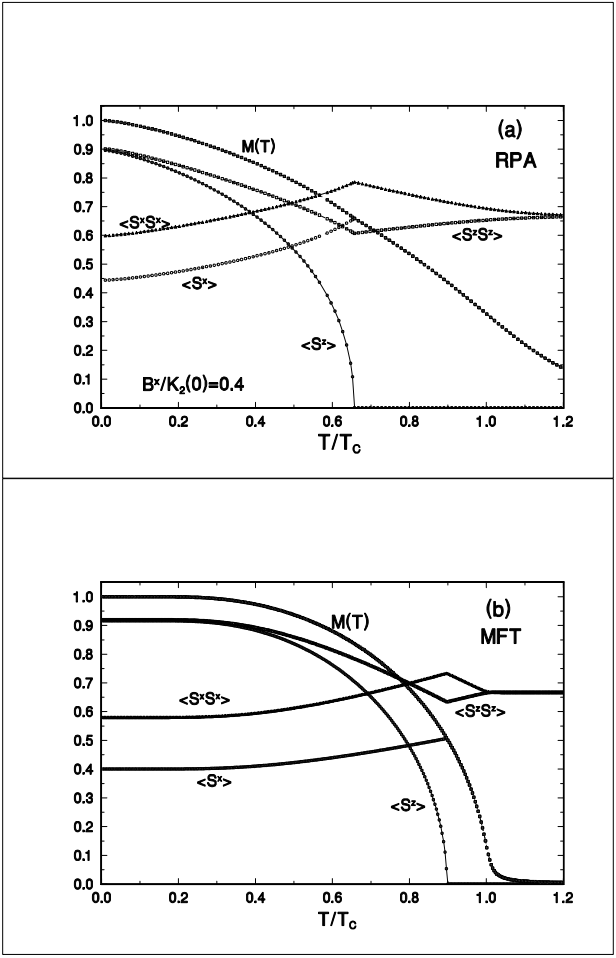}
\vspace{0cm}

\caption{
Results of RPA (a) and MFT (b) calculations for the reorientation
of the magnetization at a fixed magnetic field $B^x/K_2(0)=0.4$
are shown as functions of the reduced temperature $T/T_C$. Displayed are the
components
$\la S^z\ra$ and $\la S^x\ra$ and the magnitude $M(T)$ of the magnetization,
as well as the second moments $\la S^zS^z\ra$ and $\la S^xS^x\ra$.
}\end{figure}

\section{Discussion and Conclusion} \label{conclusion}
In the present paper, we have applied many-body Green's function theory 
for the calculation of the magnetic properties of a Heisenberg monolayer 
with second-order uniaxial single-ion anisotropy at finite temperatures.
This method allows calculations over the entire temperature
range of interest in contrast to other methods, which are only valid at low
(Holstein-Primakoff approach) or high temperatures (high temperature 
expansions). We have used the Tyablikov (RPA) decoupling for the exchange
interaction terms, and the Anderson-Callen decoupling for the
anisotr\-opy terms. For the latter, we have investigated various other 
decoupling schemes, which partly break down at higher temperatures or 
give results similar to the Anderson-Callen decoupling, see Appendix A.
The results are far more sensitive to a variation of the strength of
the anisotr\-opy than to the different
decoupling procedures. We emphasize the fact that the present method fulfills
the Mer\-min-Wagner theorem in the limiting case of an isotropic 2D Heisenberg
magnet, in contrast to the mean-field approximation applied formerly.

Our main investigations are concerned with the reorientation of the
magnetization induced by a magnetic field perpendicular to the easy
axis at finite temperatures, and with a non-perturbative calculation of the
temperature dependent (effective) anisotropy. We have in particular
investigated the monolayer for spin $S=1$ which is the lowest spin quantum
number with a non-trivial second-order anisotropy.
By solving the equations of motion for the Green's functions we  calculate
the components of the magnetization directly, which allows
an immediate determination of the orientation angle. The effective
anisotropy coefficient is calculated from the
condition that the free energy has a minimum at the orientation angle.
As discussed in Section 2 this is a non-perturbative approach in the sense
that all quantities entering in the final expression are determined by 
the \em full \em Hamiltonian. This is shown to be an improvement 
over the usual thermodynamical perturbative treatment, where
the unperturbed part of the Hamiltonian must contain the magnetic field.
Therefore the corresponding results for the effective anisotropy
necessarily depend on the magnetic field.
\begin{figure}[t] \label{Fig6}
\vspace*{-2cm}\hspace*{-0.7cm} 
\includegraphics[width=6cm,height=10cm,angle=-90]{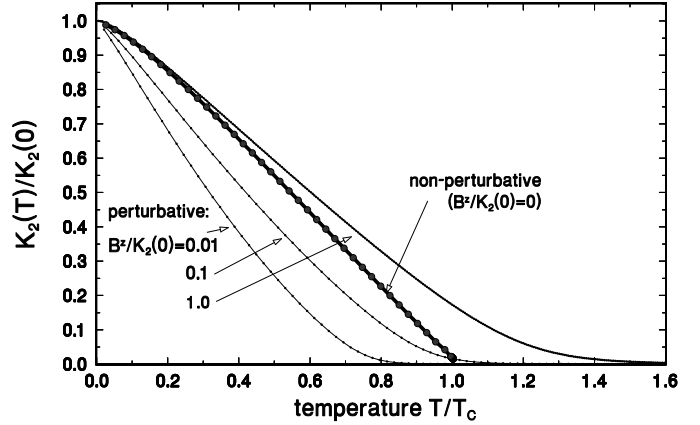}
\vspace{2cm}

\caption{Effective anisotropy coefficients $K_2(T)/K_2(0)$ calculated 
with RPA are displayed as functions of the reduced temperature $T/T_C$.
The non-perturbative RPA result is compared with perturbative RPA results 
with external fields $B^z/K_2(0)=0.01,0.1$, and $1$. }\end{figure}

Prior to the present work, the magnetic reorientation and the effective
anisotropies have been calculated mainly within the framework of mean-field
theory (MFT) \cite{JeB98,theo}. The magnetic reorientation has also been 
investigated with other theoretical methods \cite{DIPmon,MC}. In the 
presence of aniso\-tropic interactions, MFT is expected to yield 
\em qualitatively \em correct results for 2D magnetic
systems. By comparison, however, we find significant \em quantitative \em 
differences between the results obtained with MFT and the present Green's 
function theory. Owing to the magnetic fluctuations, one finds a
different temperature scale with RPA as compared to MFT. With the
same input parameters, the Curie temperature $T_C$ of the present 2D system is
about a factor of three larger in MFT than in RPA.
If the temperature is rescaled with respect to the corresponding Curie
temperatures the orientation angles of the magnetization look very similar in
RPA and MFT. This we consider to be a non-trivial result because one would not
expect this from the very beginning. The effective anisotropies, however,
behave differently, also when rescaling the temperature.
When using RPA the temperature dependence of the effective anisotropy
coefficient behaves linearly over the whole temperature range $0\leq T\leq
T_C$. In MFT one observes an exponential dependence at low temperatures 
and small $S$. This different behaviour would manifest itself when
extrapolating measurements performed at finite temperatures down to
$T=0$ in order to compare with \em ab-initio \em calculations \cite{anis} 
for the anisotropy coefficient $K_2(T=0)$, which are available only there.

One also finds in RPA a stronger influence of the external magnetic field
on the reorientation of the magnetization, which is reminiscent of 2D
Heisenberg magnets \cite{yab}. We observe that, at a fixed temperature, a
weaker reorientation field $B^x_R$  is required
in RPA than in MFT in order to align the magnetization along the field
direction. At a fixed magnetic field, on the other hand, one obtains in RPA a
lower reduced reorientation temperature $T_R/T_C$.
Furthermore, the magnetization
calculated by RPA  has a long tail at large temperatures due to the magnetic
field, which is absent in MFT results. In general, one expects the differences
between RPA and MFT to decrease as the number of film layers increases.

Investigations into extending the model are in progress. The other decoupling
procedure different from that of Anderson-Callen, which also works up
to $T_C$, see Fig.7,
will be investigated. In this case, none of the eigenvalues of the secular
problem vanishes and one has to modify the formalism appropriately. A next
step is to include the magnetic dipole coupling
competing with the uniaxial single-ion anisotropy.
The effect of the dipole coupling will become more important
when the calculations are extended to magnetic films with several layers in
order to treat also the dependence of the reorientation on the film
thickness. In this case, the layer dependent anisotropies might lead to a
temperature driven reorientation without the application of an external
magnetic field, since the temperature dependence of the anisotropies will be
different for surface and interior film layers. 
In addition, calculations for larger
spins will be attempted, at least for spin $S=2$, in order to be able to treat
the fourth-order single-ion anisotropy $K_4$. Then one can set up a phase
diagram e.g. in the $K_2-K_4$-plane, which will show the region of stable
magnetization directions and the location of the temperature driven magnetic
reorientation. Furthermore, one can investigate whether the magnetic
reorientation takes place continuously or discontinuously \cite{JeB98,theo}.

\section*{Appendix A: 
Decoupling schemes for the single-ion anisotropy term} \label{appa}
In this appendix we discuss various decoupling procedures for the
single-ion anisotropy term, and give arguments for using the
Anderson-Callen decoupling.
For an anisotropy \em strong \em compared to the exchange coupling,
a method using different Green's functions has been applied formerly
for spin $S=1$ \cite{ega}. However, this method leads to an
overdetermined system of equations for the expectation values
$\la S^z\ra$ and $\la S^zS^z\ra$. On the other hand, for
\em small \em  anisotropies,
as is the case for the magnetic systems considered in the present paper,
different approaches have been proposed by Anderson-Callen \cite{AC64} and by
Lines \cite{Lin67}, which we shall treat here.

For the decoupling of the higher-order Green's functions coming from
the single-ion anisotropy term, an ansatz of the following form is used:
\begin{equation}
\la\la S_i^zS_i^\pm+S_i^\pm S_i^z;(S_j^z)^m(S_j^-)^n\ra\ra\simeq
\Phi_i^{\pm,mn}\,G_{ij}^{\pm,mn}\;, \label{A1}
\end{equation}
where the functions $\Phi_i^{\pm,mn}$ have to be determined. For $m=0$,
$n=1$, Lines \cite{Lin67} proved that $\Phi_i^{+,01}$ can be expressed by the
following ratios of commutators
\begin{equation}
\Phi_i^{+,01}=\frac{\la[S_i^-,S_i^+(2S_i^z+1)]\ra}{\la[S_i^-,S_i^+]\ra}=
\frac{3\la(S_i^z)^2\ra-S(S+1)}{\la S_i^z\ra}\;. \label{A2}
\end{equation}
For $m=n=1$ one obtains
\bea &&\Phi_i^{+,11}=\frac{\la[S_i^zS_i^-,S_i^+(2S_i^z+1)]\ra}
{\la[S_i^zS_i^-,S_i^+]\ra}= \label{A3} \\
&&\frac{8\la(S_i^z)^3\ra-3\la(S_i^z)^2\ra+(1-4S(S+1))\la S_i^z\ra+S(S+1)}
{3\la(S_i^z)^2\ra-\la S_i^z\ra-S(S+1)}\;. \rule[-0.2cm]{0cm}{1cm}
\nonumber \eea
Note that the $\Phi_i^{+,mn}$ depend only on a single lattice site $i$,
as expected. Similar expressions may be derived for $\Phi_i^{-,mn}$.

An alternative expression for $\Phi_i^{+,mn}$ can be obtained by replacing 
the Green's functions in equation (\ref{A1}) by their respective 
expectation values, resulting in
\begin{equation}
\la (S_i^z)^m(S_i^-)^n(S_i^zS_i^\pm+S_i^\pm S_i^z)\ra\simeq
\tilde\Phi_i^{\pm,mn}\,\la (S_i^z)^m(S_i^-)^nS_i^\pm\ra\;. \label{A4}
\end{equation}
One obtains for $n=1$ equation (\ref{A5}) shown below. 
\begin{figure*}[t]
\begin{equation}
\tilde\Phi_i^{+,m1}=\frac{S(S+1)\la(S_i^z)^m\ra+
(2S(S+1)-1)\la(S_i^z)^{m+1}\ra-3\la(S_i^z)^{m+2}\ra
-2\la(S_i^z)^{m+3}\ra}{S(S+1)\la(S_i^z)^m\ra
-\la(S_i^z)^{m+1}\ra-\la(S_i^z)^{m+2}\ra}\;. \label{A5}
\end{equation} \end{figure*}
For spin $S=1$, which we treat in the present paper, we need $m=0,1$ and 
$n=1$. One has $S(S+1)=2$, $(S_i^z)^3=S_i^z$, and $(S_i^z)^4=(S_i^z)^2$, 
thus the ratios of commutators are
\begin{equation}
\Phi_i^{+,01}=\frac{3\la(S_i^z)^2\ra-2}{\la S_i^z\ra}\;,\hspace{1.3cm} 
\Phi_i^{+,11}=-1, \label{A6a} \nonumber
\end{equation}
and the ratios of the expectation values are given by
\begin{equation}
\tilde\Phi_i^{+,01}=\frac{2+\la S_i^z\ra-3\la(S_i^z)^2\ra}
{2-\la S_i^z\ra-\la(S_i^z)^2\ra}\;,\qquad \tilde\Phi_i^{+,11}=-1\,.
\label{A6} \end{equation}
Note that Lines \cite{Lin67} mixes both approaches, since he  uses
equation (\ref{A2}) for even $m$ and equation (\ref{A5}) for odd $m$.
\begin{figure}[t] \label{FigA1}
\vspace*{-0.1cm}\hspace*{-0.1cm} 
\includegraphics[width=5cm,height=8.5cm,angle=-90]{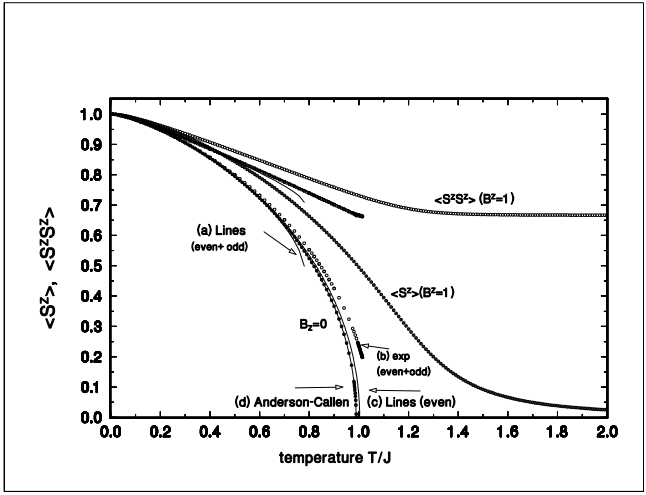}
\vspace{2cm}

\caption{The magnetization $\la S^z\ra$  and the second moment 
$\la S^zS^z\ra$ are shown as function
of the temperature for various decoupling schemes discussed in the text.
(a) decoupling of  equation (\ref{A6a}); (b) decoupling of equation
(\ref{A6}); (c) decoupling of equation (\ref{A10}); (d) Anderson-Callen 
decoupling ( equation (\ref{A12}), small dots connected by thin line).
Only the decouplings (c) and (d) work up to $\la S^z\ra=0$, i.e. up to
the Curie temperature $T_C$, whereas (a) and (b) fail for $T<T_C$.
We include also a result with Anderson-Callen decoupling and a magnetic
field $B^z/K_2(0)=1$. }\end{figure}

In order to investigate the differences between these decoupling 
procedures, we treat the special case $B^x=B^y=0$, for which
$\la S_i^x\ra=\la S_i^y\ra=0$. Then one has to determine
the expectation values $\la S_i^z\ra$ and $\la S_i^zS_i^z\ra$. For $S=1$
we obtain from equations (\ref{26b}) and (\ref{26c}),
dropping  the site index $i$ for the ferromagnetic system,
\begin{eqnarray}
\la(S^z)^2\ra&=&2-\la S^z\ra(1+2\phi_0)\,, \nonumber \\
\la S^z\ra&=&\frac{1+2\phi_0}{1+2\phi_1+\phi_0+3\phi_1\phi_0}\;,
\label{A7} \end{eqnarray}
with
\begin{eqnarray}
\phi_m&=&\frac{1}{N}\sum_{\bf k}\bigg(\exp(\beta E_{\bf k}^m)-1\bigg)^{-1}
\label{A8}\,, \\
E_{\bf k}^m&=&B^z+\la S^z\ra J(4-\gamma_{\bf k})+K_2\,\Phi^m \;,
\label{A9} \end{eqnarray}
with $m=0,1$, where $\Phi^m$ is either $\Phi^{+,m1}$ or $\tilde\Phi^{+,m1}$.

When solving these equations for the decoupling procedures
(\ref{A6a}), (\ref{A6}), we find that both procedures fail at 
temperatures far below $T_C$, cf.\ Fig.7 (situation (a),(b)).
Thus one cannot use both decouplings up to the Curie temperature $T_C$.

When calculating Curie temperatures, Lines \cite{Lin67} has
circumvented the problem associated with equations (\ref{A7}) by
(inconsistently) using in the \em dispersion relations \em $E_{\bf k}^m$,
equation (\ref{A9}), expectation values $\la S^z\ra$ and $\la S^zS^z\ra$ as
obtained from the one-particle density operator, using a theorem originating
from Callen and Strikman \cite{CaS65}.

If, however, one uses
\begin{equation}
\Phi^{+,m1}=\Phi^{+,01}\  ({\rm or\ } \tilde\Phi^{+,m1}=\tilde\Phi^{+,01},
{\rm respectively}) \label{A10}
\end{equation}
for $m=0,1$, the procedure works well and yields reasonably results
for $\la S^z\ra$ and $\la S^zS^z\ra$ between $T=0$ and $T=T_C$, cf.\
Fig.7 (situation (c)), and the following discussion about 
the Curie temperature $T_C$.

In this case, $\phi_0=\phi_1\equiv\phi$ , see equation (\ref{A8}), and one
finds
that the procedures for $\Phi^{+,01}$ and $\tilde\Phi^{+,01}$ give the same
result. This can be understood by equating $\Phi^{+,01}=\tilde\Phi^{+,01}$,
which yields the second of equations (\ref{A7}) determining $\la S^z\ra$.

Using this decoupling in the Green's function formalism one can derive
expressions for the $\Phi^{-,mn}$, using also $\Phi^{-,01}=\Phi^{-,11}$.
This leads to a $3\times 3$ secular problem for equation (\ref{6})
with three non-vanishing eigenvalues, since here 
$\Phi^{+,mn}\neq \Phi^{-,mn}$. In this case the regularity condition 
(\ref{18}) cannot be applied. This case is more complicated to handle 
than the Anderson-Callen decoupling, in which one of the eigenvalues 
vanishes, cf.\ equation (\ref{11}).

The Anderson-Callen decoupling is based on the paper by Callen \cite{Cal63}, 
in which correlations beyond the RPA are included in the decoupling of the
Green's functions of the \em exchange \em terms. In this case, the
essential Green's function is non-diagonal ($i\neq j$) and the
decoupling reads (e.g.\ for $m=0$, $n=1$)
\begin{equation}
\la\la S_i^zS_j^\pm;S_k^-\ra\ra\simeq \la S_i^z\ra
\la\la S_j^\pm;S_k^-\ra\ra-\frac{\la S_i^z\ra}{2S^2}\la S_i^\mp 
S_j^\pm\ra\la\la S_i^\pm;S_k^-\ra\ra\,. \label{A11}
\end{equation}
Neglecting the second term corresponds to the RPA decoupling.

The proposal of Anderson and Callen \cite{AC64} is to use the same decoupling
also for Green's functions with $i=j$, resulting from the anisotropy terms
\bea &&\la\la(S_i^zS_i^\pm+S_i^\pm S_i^z);S_k^-\ra\ra \nonumber \\
&\simeq&2\la S_i^z\ra\Big\{1-\frac{1}{4S^2}\Big[\la S_i^\mp S_i^\pm\ra
+\la S_i^\pm S_i^\mp\ra\Big] \Big\}\la\la S_i^\pm;S_k^-\ra\ra \nonumber \\
&=&2\la S_i^z\ra\Big\{1-\frac{1}{2S^2}\Big[S(S+1)- \la S_i^z
S_i^z\ra\Big]\Big\}\la\la S_i^\pm;S_k^-\ra\ra \nonumber \\
&=&\Phi^\pm\la\la S_i^\pm;S_k^-\ra\ra\,. \label{A12} \eea
In this case, one does not distinguish between different $m$. Since here
$\Phi^+=\Phi^-=\Phi$, which is the quantity $\Phi$ given in equation
(\ref{8}), one of the eigenvalues turns out to be zero, cf.\ equation 
(\ref{11}). This is a prerequisite for being able to use the procedure
outlined in Section 2 for
the determination of the expectation values $\la S^z\ra$ and $\la S^zS^z\ra$.

Inspecting Fig.7 shows that the two decouplings, \break which 
work up to the Curie temperature, do not give  very different results.
Therefore, we adopt the decoupling which is easier to apply:
the Anderson-Callen decoupling.

In the remainder of this appendix, we derive expressions which determine the
Curie temperature, $T_C(J,K_2)$,  for the case $S=1$ and
$\Phi^{+,01}=\Phi^{+,11}=\Phi$, cf.\ equation (\ref{A10}).
The consideration of different $\Phi^{+,01}\neq\Phi^{+,11}$,
cf.\ equation (\ref{A6a}), leads to severe complications
whilst determining $T_C$, or, as we have seen, when calculating the 
magnetization $\la S^z\ra$ as function of $T$.
\begin{figure}[t] \label{FigA2}
\vspace*{-1.6cm}\hspace*{-0.7cm} 
\includegraphics[width=6cm,height=10cm,angle=-90]{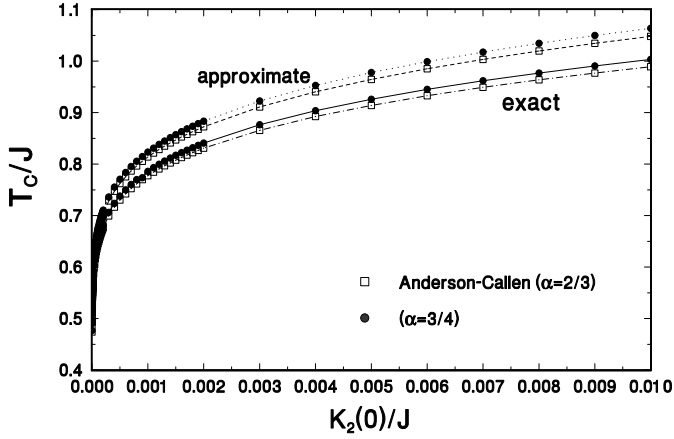}
\vspace{2cm}

\caption{
Curie temperatures normalized to the exchange coupling, $T_C/J$, are shown as
a function of the anisotropy coupling strength $K_2(0)/J$,
using equation (\ref{A17}) with $\alpha=3/4$ (decoupling according to equation
(\ref{A10})) and $\alpha=2/3$ corresponding to the Anderson-Callen decoupling
(equation (\ref{A12}). The Curie temperatures at $K_2(0)/J=0.01$ correspond to
those of Fig.7 (cases (c) and (d)). Also, the results from the
approximate expression, equation (\ref{A18}), for $\alpha=3/4$ and 
$\alpha=2/3$  are included in the figure. }\end{figure}

$T_C$ is calculated from performing the limit
$\la S^z\ra\to0$ in equations (\ref{A7}).
To lowest order in $\la S^z\ra$, one obtains for $T\to T_C$
\begin{equation}
\la (S^z)^2\ra(T\ltsim T_C)\approx 2/3+\la S^z\ra^2/4\,,
\label{A13} \end{equation}
and for the decoupling functions 
\begin{equation}
\Phi(T_C)=\alpha\; \la S^z\ra\,, \label{A14}
\end{equation}
with $\alpha=3/4$ for the decoupling given by equation (\ref{A10}) and
$\alpha=2/3$ for the Anderson-Callen decoupling, equations (\ref{8}) or
(\ref{A12}). Close to $T_C$ one also obtains from equation (\ref{A8})
\begin{equation}
\phi(T_C)\approx \frac{T_C}{\la S^z\ra}\;\frac{1}{N}\sum_{\bf k}
\Big(J(4-2\gamma_{\bf k})+K_2\,\alpha\Big)^{-1}\,, \label{A15}
\end{equation}
and from equation (\ref{A7})
\begin{equation}
\phi(T_C)\approx \frac{2}{3\la S^z\ra}\,.
\label{A16} \end{equation}
Combining these equations and converting the sum into an integral gives
\begin{equation}
T_C=\frac{8\pi^2}{3}\bigg(  \int_{-\pi}^\pi dk_x\int_{-\pi}^\pi dk_y
\frac{1}{J(4-\gamma_{\bf k})+K_2\,\alpha}\bigg)^{-1}\,.
\label{A17} \end{equation}
The Curie temperatures calculated with this formula are displayed in
Fig.8 for the two single-ion decouplings under consideration.

Since this integral is dominated by small wave numbers
it can be approximately evaluated when
expanding $\gamma_{\bf k}$ up to the leading order in {\bf k}:
\begin{equation}
T_C=\frac{8\pi J/3}{\ln(1+2\pi^2J/K_2\,\alpha)}\,.
\label{A18} \end{equation}
This expression can be used as a quick estimate for the Curie temperature. It
overestimates the result as obtained from equation (\ref{A17}) by less than
$10\%$. The corresponding results are also shown in Fig.8.
\begin{figure}[t] \label{FigA3}
\vspace*{-2cm}\hspace*{2cm} 
\includegraphics[width=9cm,height=13cm,angle=180]{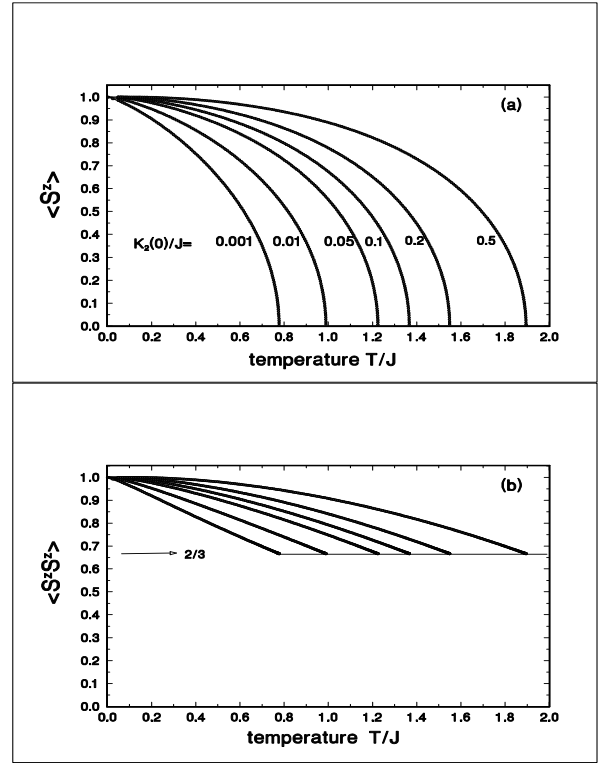}
\caption{
The temperature dependence of $\la S^z\ra$ (a) and $\la S^zS^z\ra$ (b)
for different strengths of the anisotropy coefficient
$K_2(0)/J=0.001, 0.01, 0.05, 0.1, 0.2,$ and $0.5$ is displayed.
}\end{figure}

\renewcommand{\arraystretch}{1.5}
\begin{table*}[t] \label{T1}
\begin{tabular}{|l|l|l|l|l|l|l|}  \hline
m & n &$A_{-1}^{z,mn}$ & $A_{-1}^{+,mn}$& $A_{-1}^{-,mn}$ \\ \hline
0 & 1 & $-\la S^-\ra$  & $2\la S^z\ra$  & 0 \\
0 & 2 & $-2\la S^-S^-\ra$ & $4\la S^zS^-\ra+2\la S^-\ra$ & 0 \\
1 & 1 & $-\la S^zS^-\ra$ & $3\la S^zS^z\ra -\la S^z\ra-S(S+1)$
  & $\la S^-S^-\ra$ \\
0 & 3 & $-3\la S^-S^-S^-\ra$ & $6\la S^zS^-S^-\ra+6\la S^-S^-\ra$ & 0 \\ 
1 & 2 & $-2\la S^zS^-S^-\ra$  & $5\la S^zS^zS^-\ra
  +\la S^zS^-\ra-\la S^-\ra S(S+1)$&$\la S^-S^-S^-\ra$\\
2 & 1 &  $-\la S^zS^zS^-\ra$ & $4\la S^zS^zS^z\ra-3\la S^zS^z\ra $
  & $2\la S^zS^-S^-\ra+\la S^-S^-\ra$ \\
  & & & $+(1-2S(S+1))\la S^z\ra+S(S+1)$& \\ \hline
\end{tabular}
\vspace{0.5cm}

\parbox{13cm}{\caption{Inhomogeneities 
$A_{-1}^{\alpha,mn}=\la[S^\alpha,(S^z)^m(S^-)^n]_{-1}\ra$, $\alpha=+,-,z$,  
of the anticommutator Green's functions, cf.\ equation (\ref{2}), 
for different integers $m$ and $n$. }}
\end{table*}
\renewcommand{\arraystretch}{1.2}

\setcounter{equation}{56}
\begin{figure*}[t]
\bea \la S^zS^zS^-\ra&=&\frac{B^-Z}{2Z^2-3B^+B^-}\Big[4\la
S^zS^zS^z\ra-\la S^z\ra\Big(2S(S+1)-1\Big)\nonumber\\
& &-\Big(3\la S^zS^z\ra-S(S+1)\Big)\frac{2Z^2-3B^+B^-}{2Z^2-B^+B^-}
-\frac{B^+B^-}{Z^2}\Big(\la S^zS^zS^z\ra+\frac{1}{2}\la
S^z\ra\Big)\Big] \nonumber\\
\la S^zS^-S^-\ra&=&\frac{(B^-)^2}{2Z^2-3B^+B^-}\Big[5\la
S^zS^zS^z\ra-\la S^z\ra\Big(3S(S+1)-1\Big)-\Big(3\la
S^zS^z\ra-S(S+1)\Big)\frac{2Z^2-3B^+B^-}{2Z^2-B^+B^-}\Big] \nonumber\\
\la S^-S^-S^-\ra&=& \frac{(B^-)^3}{Z(2Z^2-3B^+B^-)}\Big[5\la
S^zS^zS^z\ra-\la S^z\ra\Big(3S(S+1)-1\Big)\Big] \,. \label{B3}
\eea \end{figure*}
\setcounter{equation}{53}

We emphasize the fact that the various decoupling procedures do not yield very
different results for the magnetization, see Fig.7, or the Curie
temperature $T_C$, cf.\ Fig.8. We note that the results are much 
more sensitive to varying the strength of the anisotropy coefficient $K_2(0)$
than to the different decoupling procedures. This is shown in Fig.9 
by plotting the magnetization $\la S^z\ra$ and $\la S^zS^z\ra$
as function of the temperature for different values $K_2(0)/J$ of the
anisotropy coefficient normalized to the exchange coupling strength $J$
for the case of the Anderson-Callen decoupling.

\section*{Appendix B} \label(appb)
In this Appendix we list explicitly a number of relations obtained from the
regularity condition equation (\ref{17})
\begin{equation}
-2ZA_{-1}^{z,mn}=A_{-1}^{+,mn}B^-+A_{-1}^{-,mn}B^+ \,.
\end{equation}
Remember that we have used the notations $B^\pm=B^x\pm i\,B^y$,
$Z=B^z+K_2\,\Phi$,
and $A_{-1}^{\alpha,mn}=\la[S^\alpha,(S^z)^m(S^-)^n]_{-1}\ra$.
In Table \ref{T1} we tabulate some commutator relations. 

We obtain with $m=0, n=1$
\begin{equation}
\la S^\pm\ra =\frac{B^\pm}{Z}\la S^z\ra \,, \label{B1}
\end{equation}
with $m=0, n=2$ and $m=1, n=1$
\begin{eqnarray}
\la S^-S^-\ra&=&\frac{(B^-)^2}{2Z^2-B^+B^-}\Big(3\la S^zS^z\ra-S(S+1)\Big)
\nonumber \\
\la S^zS^-\ra&=&\frac{B^-Z}{2Z^2-B^+B^-}\Big(3\la S^zS^z\ra-S(S+1)\Big) 
\nonumber \\
&& -\frac{B^-}{2Z}\la S^z\ra \,, \label{B2}
\end{eqnarray}
and with $m=0, n=3$, and $m=1, n=2$, and $m=2, n=1$
given in equations (\ref{B3}) displayed above.


\begin{thebibliography}{99}
\bibitem{Hein} J.A.C. Bland, B. Heinrich, \em Ultrathin Magnetic Structures,
\em (Springer Verlag, Berlin, 1994).
\bibitem{exp} G. Lugert, W. Robl, L. Pfau, M. Brockmann,
G. Bayreuther, J. Magn. Magn. Mater. {\bf121}, 498 (1993);
O. Schulte, F. Klose, W. Felsch, Phys. Rev. B {\bf52},
6480 (1995); M. Farle, B. Mirwald-Schulz, A.N. Anisimov, W. Platow,
K. Baberschke, Phys. Rev. B {\bf55}, 3708 (1997), and references therein.
\bibitem{dut} J.R. Dutcher, J.F. Cochran, I. Jacob, W.F. Egelhoff, Jr.,
Phys. Rev. B {\bf 39}, 10430 (1989).
\bibitem{MOK98} Y.T. Millev, H.P. Oepen, J. Kirschner,
Phys. Rev. B {\bf57},  5837, 5848 (1998);
 A. Hucht and K.D. Usadel,  preprint cond-mat/9903040.
\bibitem{anis} O. Hjortstam, K. Baberschke, J.M. Wills, B. Johansson,
O. Erickson, Phys. Rev. B {\bf55}, 15 026 (1997);C. Uiberacker, J. Zabloudil,
P. Weinberger, L. Szunyogh, C. Sommers,
Phys. Rev. Lett. {\bf82}, 1289 (1999), and references therein.
\bibitem{JeB98} A. Moschel, K.D. Usadel, Phys. Rev. B {\bf 49}, 12868 (1994);
P.J. Jensen, K.H. Bennemann, {\it in `Magnetism and Electronic Correlations
in Local-Moment Systems: Rare-Earth Elements and Compounds'},
ed. M. Donath, P.A. Dowben and W. Nolting, (World Scientific, Singapore, 1998),
p. 113-141.
\bibitem{theo} P.J. Jensen, K.H. Bennemann, Solid State Comm. {\bf100}, 
585 (1996); \em ibid. \em {\bf105}, 577 (1998);
A. Hucht, K.D. Usadel, Phys. Rev. B {\bf55}, 12309 (1997).
\bibitem{MW66} N.M. Mermin, H. Wagner, Phys. Rev. Lett. {\bf17}, 1133 (1966).
\bibitem{herr} C. Herring, C. Kittel, Phys. Rev. {\bf 81}, 869 (1951);
S.V. Maleev, Sov. Phys. JETP {\bf 43}, 1240 (1976); V.L. Pokrovsky, M.V.
Feigel'man, \em ibid., \em {\bf 45}, 291 (1977).
\bibitem{yab} D.A. Yablonsky, Phys. Rev. B {\bf 44}, 4467 (1991);
D. Kerkmann, D. Pescia, R. Allenspach, Phys. Rev. Lett. {\bf 68}, 686 (1992).
\bibitem{Mills} R.P. Erickson, D.L. Mills, Phys. Rev. B {\bf 43}, 10 715
(1991); \em ibid. \em {\bf44}, 11 825 (1991).
\bibitem{HP40} T. Holstein, H. Primakoff, Phys. Rev. {\bf58}, 1098 (1940).
\bibitem{EFJK99} A. Ecker, P. Fr\"obrich, P.J. Jensen, P.J. Kuntz,
J. Phys.: Condens. Matter {\bf 11}, 1557 (1999).
\bibitem{Tya67} S.V. Tyablikov, \em Methods in the quantum theory of
magnetism \em (Plenum Press, New York, 1967);
K. Elk and W. Gasser, \em Die Methode der Greenschen Funktionen in der
Festk\"orperphysik \em (Akademie-Verlag, Berlin, 1978);
W. Nolting, \em Quantentheorie des Magnetismus, \em vol.2 (B.G.
Teubner, Stuttgart, 1986).
\bibitem{TGH98} C. Timm, S.M. Girvin, P. Henelius, A.W. Sandvik,
Phys. Rev. B {\bf 58}, 1464 (1998).
\bibitem{DIPmon} M.M. Taylor, B.L. Gy\"orffy, J. Phys.: Condens. Matter {\bf
5}, 4527 (1993); A. Moschel, K.D. Usadel, Phys. Rev. B {\bf51}, 16111 (1995);
A. Hucht, A. Moschel, K.D. Usadel, J. Magn. Magn. Mater. {\bf148}, 32 (1995).
\bibitem{AC64} F.B. Anderson, H.B. Callen,  Phys. Rev. {\bf136}, A1068 (1964).
\bibitem{MF95} Y. Millev, M. F\"ahnle, Phys. Rev. B {\bf32}, 4336 (1995).
\bibitem{MHB96} T.H. Moos, W. H\"ubner, K.H. Bennemann, Solid State Commun.
{\bf 98}, 639 (1996).
\bibitem{Tya59} S.V. Tyablikov, Ukr. Mat. Zh.{\bf11}, 287 (1959).
\bibitem{Na65} A. Narath,  Phys. Rev. {\bf 140}, A854 (1965).
\bibitem{Lin67} M.E. Lines, Phys. Rev. {\bf156}, 534 (1967).
\bibitem{MC} P. Politi, A. Rettori, M.G. Pini, D. Pescia,
J. Magn. Magn. Mater. {\bf140-144}, 647 (1995);
X. Hu, Y. Kawazoe, Phys. Rev. B {\bf54}, 65 (1996);
T. Herrmann, M. Potthoff, W. Nolting, Phys. Rev. B {\bf 58}, 831 (1998).
\bibitem{ega} T. Egami, M.S.S. Brooks, Phys. Rev. B {\bf 12}, 1021, 1029
(1975); S.B. Haley, \em ibid., \em {\bf 17}, 337 (1978).
\bibitem{CaS65} H.B. Callen, S.Strickman, Solid State Commun. {\bf3}, 5 (1965).
\bibitem{Cal63} H.B. Callen,  Phys. Rev. {\bf 130}, 890 (1963).
\end{thebibliography}
\end{document}